\documentclass[pra,twocolumn,showpacs,amsmath,amssymb]{revtex4}
\usepackage[latin2]{inputenc}
\usepackage{amsmath}
\usepackage{graphicx}
\usepackage{dcolumn}
\usepackage{epstopdf}

\newcommand{\ba}{\begin{eqnarray*}}
\newcommand{\ea}{\end{eqnarray*}}
\newcommand{\baa}{\begin{eqnarray}}
\newcommand{\eaa}{\end{eqnarray}}
\def\bar{\begin{array}}
\def\ear{\end{array}}
\def\LB{\left(}
\def\RB{\right)}

\def\pr{^{\prime}}

\def\u{\uparrow}
\def\d{\downarrow}

\def\s{\sigma}
\def\f{\frac}

\begin{document}

\title{Time-dependent occupation numbers in reduced-density-matrix functional theory: Application to an interacting Landau-Zener model}

\author{Ryan Requist\footnote{Present address: SISSA, via Bonomea 265, Trieste 34151, Italy}}
\author{Oleg Pankratov}
\affiliation{
Theoretische Festk\"orperphysik, Universit\"at Erlangen-N\"urnberg, Staudtstra\ss e 7-B2, 91058 Erlangen, Germany
}

\date{\today}

\begin{abstract}

We prove that if the two-body terms in the equation of motion for the one-body reduced density matrix are approximated by ground-state functionals, the eigenvalues of the one-body reduced density matrix (occupation numbers) remain constant in time.  This deficiency is related to the inability of such an approximation to account for relative phases in the two-body reduced density matrix.  We derive an exact differential equation giving the functional dependence of these phases in an interacting Landau-Zener model and study their behavior in short- and long-time regimes.  The phases undergo resonances whenever the occupation numbers approach the boundaries of the interval $[0,1]$.  In the long-time regime, the occupation numbers display correlation-induced oscillations and the memory dependence of the functionals assumes a simple form.

\end{abstract}


\maketitle

\section{\label{sec:intro} Introduction}

The effective single-particle Schr\"odinger equations in reduced-density-matrix functional theory \cite{gilbert1975} (RDMFT) differ from the Kohn-Sham (KS) \cite{kohn1965} and Hartree-Fock (HF) equations in that essentially all orbitals have fractional \mbox{($0<n_i<1$)} as opposed to integer occupation numbers.  The freedom to occupy orbitals fractionally is helpful in describing strongly correlated systems, where strong quantum fluctuations among configurations can cause the orbitals to have average occupations that differ significantly from $0$ or $1$.  The KS and HF equations attempt to reproduce certain observables, the density and energy, respectively, with only a single configuration (Slater determinant).  As a consequence, the orbitals can lose any resemblance to the optimal orbitals for describing the wave function, the so-called \mbox{\textit{natural orbitals} \cite{loewdin1955}.}  In dynamical problems, time-dependent occupation numbers represent changes in the degree of correlation \cite{appel2007,appel2010}.  The linear response of the occupation numbers has been shown to be crucial for describing double excitations in linear response theory \cite{giesbertz2008,giesbertz2009}.  

In RDMFT, the wave function is interpreted as a functional of the one-body reduced density matrix (one-matrix) $\gamma(1,1\pr,t) = \langle \Psi(t) | \hat{\psi}^{\dag}(1\pr) \hat{\psi}(1) | \Psi(t) \rangle$ ($1=\mathbf{r}_1,\sigma_1$).  The equation of motion is (in units $|e| = \hbar = m = c = 1$)
\begin{align}
i\partial_t \hat{\gamma} = \big[\f{1}{2} (\hat{\mathbf{p}}-\hat{\mathbf{A}})^2 + \hat{v}, \hat{\gamma} \big] + i \hat{u}, \label{eqn:EOM}
\end{align}
where $\hat{v}$ and $\hat{\mathbf{A}}$ describe time-dependent external electromagnetic fields. In the spatial representation, $\hat{u}$ is
\begin{align}
\langle 1 | \hat{u} | 1\pr \rangle = \frac{2}{i}\int d2 \left[ v_C(1,2)-v_C(1\pr,2) \right]\,\Gamma(12,1\pr2,t).
\end{align}
Here, $\Gamma(12,1\pr2\pr,t) = \frac{1}{2} \langle \Psi(t) |\hat{\psi}^{\dag}(1\pr) \hat{\psi}^{\dag}(2\pr) \hat{\psi}(2) \hat{\psi}(1) | \Psi(t) \rangle$ is the two-body reduced density matrix (two-matrix) and $v_C$ is the Coulomb potential.  Equation~(\ref{eqn:EOM}) can be closed by interpreting the two-matrix as a functional of the one-matrix and the initial state.  In fact, it follows \cite{pernal2007a,rajam2009} from the Runge-Gross theorem \cite{runge1984}, or its extension \cite{ghosh1988,vignale2004}, that there exists an exact two-matrix functional $\Gamma([\gamma],t)$, where we have suppressed the initial state dependence.  In general $\Gamma([\gamma],t)$ is a memory-dependent functional; i.e., it depends on $\gamma(t\pr)$ for all $t\pr \leq t$.  Instead of propagating the one-matrix directly, it may be more convenient to propagate its eigenfunctions and eigenvalues, called natural orbitals and occupation numbers, respectively, according to the equations \cite{pernal2007a,appel2007,requist2010a}
\begin{align}
i \,| \dot{\phi}_k \rangle &=  \big[ \frac{1}{2} (\hat{\mathbf{p}}-\hat{\mathbf{A}})^2 + \hat{v} + \hat{v}_{ee} \big] \left| \phi_k \right>, \label{eqn:orb} \\[0.2cm]
\dot{n}_k &= \left< \phi_k \left| \hat{u} \right| \phi_k \right> \nonumber \\
&= 4\: \mathrm{Im} \sum_{ijl} \Gamma_{ijkl} V_{klij} \label{eqn:occnum}
\end{align}
where the dot represents the time derivative, $v_{ee,jk}=\langle \phi_j | \hat{v}_{ee} | \phi_k \rangle =i u_{jk}/(n_k-n_j)$ ($j\neq k$), and $\Gamma_{ijkl}$ and $V_{ijkl}$ are the two-matrix and Coulomb integral expressed in the natural orbital basis, respectively.

In this paper, we report four fundamental results concerning the time dependence of the occupation numbers: (i) a proof that the occupation numbers are time independent when the exact functional $\Gamma([\gamma],t)$ is approximated by the adiabatic extension of \textit{any} ground-state (gs) functional $\Gamma[\gamma]$, (ii) an explicit differential equation for the exact memory-dependent functional $\Gamma([\gamma],t)$ in an interacting generalization of the Landau-Zener (LZ) model, (iii) the identification and characterization of correlation-induced oscillations in the occupation numbers, and (iv) the identification of a universal resonance phenomenon responsible for maintaining the Pauli exclusion principle in real-time dynamics.  Result (i) establishes the need for memory-dependent approximations to $\Gamma([\gamma],t)$, while (ii-iii) provide exact formulas for memory dependence in an important generic case, namely, the crossing of two single-particle levels coupled by interactions.

\section{\label{sec:AEA} Deficiency of the adiabatic extension approximation}

If the external potentials $v(\mathbf{r},t)$ and $\mathbf{A}(\mathbf{r},t)$ are slowly changing functions of time, the wave function remains close to the instantaneous ground state.  Therefore, it is natural to approximate $\Gamma([\gamma],t)$ in Eqs.~(\ref{eqn:orb}) and (\ref{eqn:occnum}) by the gs functional $\Gamma[\gamma]$, which we refer to as an adiabatic extension approximation (AEA).  However, the AEA was found to have the deficiency that if a HF-type gs functional is used, the occupation numbers remain constant in time \cite{pernal2007b,appel2007,appel2010}.  Since this result relied on the special form of HF-type functionals, it left open the possibility that time-dependent (td) occupation numbers could be obtained with more general gs functionals.  Recently, it was stated that the occupation numbers are \textit{always} constant in the AEA \cite{giesbertz2010b}, regardless of the gs functional that is used.  It is important to know whether this statement is true because it has implications for the design of functionals capable of changing the occupation numbers.  Although the arguments given in Ref.~\onlinecite{giesbertz2010b} to support the statement are incorrect \cite{requist2010b}, the statement is indeed true.  We now present a simple proof.

Consider a system in its ground state at $t=t_0$ that experiences the external driving $v(\mathbf{r},t)$ and $\mathbf{A}(\mathbf{r},t)$ for $t\geq t_0$.  If we exclude external driving that turns on discontinuously, then $\dot{\hat{\gamma}}(t_0) = 0$ because the system is in a stationary state at $t=t_0$.  This implies $\dot{n}_k(t_0)=0$.  Since the AEA is based only on the gs functional $\Gamma[\gamma]$, at any later time the right-hand side (rhs) of Eq.~(\ref{eqn:occnum}) in the AEA is the same as the \textit{exact} rhs we would have for a system just starting in its gs at that time.  Thus, it vanishes because the gs is a stationary state.  We have assumed that the $\gamma(t)$ obtained in the AEA remains gs ensemble $v$-representable by a local or nonlocal external potential $v(\mathbf{r}\sigma,\mathbf{r}\pr \sigma\pr)$.  If it did not, $\Gamma[\gamma]$ on the rhs of Eq.~(\ref{eqn:occnum}) would become ill-defined.  

The natural orbitals are not constant in the AEA since they are driven by the external fields.  The arguments used in the proof do not apply to Eq.~(\ref{eqn:orb}) because its rhs cannot be interpreted as a gs functional.  There is a mismatch due to the presence of $v(\mathbf{r},t)$ and $\mathbf{A}(\mathbf{r},t)$.  Namely, the gs $|\Psi\rangle$ uniquely determined \cite{gilbert1975} by the instantaneous $\gamma(t)$ is generally not the instantaneous gs $|\Psi\rangle$ corresponding to $v(\mathbf{r},t)$ and $\mathbf{A}(\mathbf{r},t)$.

In contrast, the occupation numbers are not driven directly by the external fields because $v(\mathbf{r},t)$ and $\mathbf{A}(\mathbf{r},t)$ do not appear in Eq.~(\ref{eqn:occnum}).  Instead, they are driven \textit{purely} by the internal correlation of the system, as only the correlation part of the two-matrix gives a nonvanishing contribution on the rhs of Eq.~(\ref{eqn:occnum}) \cite{appel2007,appel2010}.  The correlation part of the two-matrix is defined as $\Gamma_{\rm c} = \Gamma - \Gamma_{HF}$, where $\Gamma_{HF}(11\pr,22\pr) = \gamma(1,1\pr) \gamma(2,2\pr) - \gamma(1,2\pr) \gamma(2,1\pr)$ is the HF two-matrix.  On the basis of the above proof, we can make the stronger statement that the $n_k$ are driven purely by \textit{nonadiabatic correlation}, i.e., the difference between $\Gamma_{\rm c}$ and its instantaneous gs value.  

What is $\Gamma[\gamma]$ missing that makes it incapable of generating $\dot{n}_k \neq 0$ in Eq.~(\ref{eqn:occnum})?  For two-electron singlet states, the only difference between the exact functional $\Gamma([\gamma],t)$ and $\Gamma[\gamma]$ are relative phases that correspond to the relative phases between the configurations that compose the wave function \cite{requist2010a}.  These relative two-matrix phases must differ from their gs values to yield $\dot{n}_k \neq 0$ in Eq.~(\ref{eqn:occnum}).  The AEA fails because, being based solely on the gs functional $\Gamma[\gamma]$, it cannot change the two-matrix phases away from their gs values.  In the general $N$-electron case, the functional $\Gamma([\gamma],t)$ differs from $\Gamma[\gamma]$ in more degrees of freedom than just these relative phases.  An alternative functional theory that might allow one to take into account the relative two-matrix phases in an effective way has been introduced \cite{giesbertz2010b}.  The phases of the natural orbitals, which in RDMFT are undefined, are incorporated into the basic independent variable of the functional.  

The relative phases can be seen explicitly by considering the L\"owdin-Shull \cite{loewdin1956} wave function for two-electron singlet states, which can be written as
\begin{align}
|\Psi^{LS}\rangle = \f{1}{\sqrt{2}} e^{-i\mu} \sum_k e^{-i2\zeta_k} \sqrt{n_k} \: \hat{a}_{k\uparrow}^{\dag} \hat{a}_{k\downarrow}^{\dag}|\rangle, \label{eqn:psi:LS}
\end{align}
where $\hat{a}_{k\sigma}^{\dag}$ are the natural orbital creation operators and $\zeta_k$ are the relative phases between the configurations.  The corresponding expression for the two-matrix in the basis of natural orbitals, $\Gamma_{ijkl} = \f{1}{2} e^{i2(\zeta_k-\zeta_i)} \sqrt{n_i n_k} \delta_{ij}\delta_{kl}$, shows that varying the $\zeta_k$ changes the phases of the elements $\Gamma_{ijkl}$, i.e., it changes relative phases in the two-matrix.  There is a one-to-one correspondence between the relative two-matrix phases and the $\zeta_k$.  The $\zeta_k$ are not invariant under redefinitions of the natural orbital phases, so they can only be uniquely defined with respect to a given choice of td natural orbital phases.  However, for any choice of td natural orbital phases, the $\zeta_k$ have unique gs values.  To generate td occupation numbers in Eq.~(\ref{eqn:occnum}), the $\zeta_k$ in the functional $\Gamma([\gamma],t)$ must differ from their gs values.  Memory dependence enters $\Gamma([\gamma],t)$ exclusively through the $\zeta_k$, which are functionals of $\gamma(t)$.  The relative phases can be given a geometric interpretation, which will be discussed in Sec.~\ref{sec:model}. 
  
Despite the failure of the AEA in Eq.~(\ref{eqn:occnum}), it is possible to obtain td occupation numbers in an approach based only on the gs functional $\Gamma[\gamma]$.  In the adiabatic regime, td occupation numbers can be obtained on the fly during the propagation of Eq.~(\ref{eqn:orb}) from a condition of instantaneous occupation number relaxation (IONR) \cite{requist2010a}.  The IONR approximation reproduces an adiabatic approximation in linear response theory \cite{pernal2007b}, but it is more general because it applies to fully nonlinear real-time dynamics.  It captures the lowest-order non\-adiabatic effects even though it lacks memory dependence.  However, if the occupation numbers deviate greatly from their instantaneous gs values, the IONR approximation breaks down.  In such cases, it is necessary to propagate Eq.~(\ref{eqn:occnum}).  Recently, an approach was proposed in which an approximation for $\Gamma_{\rm c}$ is obtained from semiclassical propagation of the $N$-body density matrix.  This $\Gamma_{\rm c}$ is then used in the propagation of Eq.~\ref{eqn:EOM}, yielding td occupation numbers.

\section{\label{sec:model} Interacting Landau-Zener model}

Consider two generic single-particle states whose bare energies cross linearly in time.  If we occupy the system with two electrons and allow interactions between them, we obtain an interacting generalization of the Landau-Zener (LZ) model.  We consider the dynamics in the sector of spin-singlet states.  The extension to the full Hilbert space is an interesting problem for future study.  The spin-singlet sector has three states.  Having at least three states is essential for representing genuine interactions.  The three levels undergo a \textit{correlated} avoided crossing (see Fig.~4 in Ref.~\onlinecite{requist2010a}), which is one of a hierarchy of multiplet-like avoided crossings that one encounters in many-body systems.  Occupation numbers vary most rapidly near such avoided crossings.  The Hamiltonian with the most general one-body and two-body terms in the spin-singlet sector is 
\begin{align}
\hat{H} &= \frac{1}{2} \vec{V}\cdot\hat{\vec{\sigma}} + \hat{U} + \hat{W},
\end{align}
where $\hat{\vec{\sigma}} = \sum_{\s} (\hat{c}_{1\s}^{\dag},\: \hat{c}_{2\s}^{\dag}) \vec{\sigma} \LB \bar{c} \hat{c}_{1\s} \\ \hat{c}_{2\s} \ear \RB$, $\vec{V}$ acts as an external potential, and
\begin{align}
\hat{U} &=  U_1 \hat{c}_{1\uparrow}^{\dag} \hat{c}_{1\uparrow} \hat{c}_{1\downarrow}^{\dag} \hat{c}_{1\downarrow} + U_2 \hat{c}_{2\uparrow}^{\dag} \hat{c}_{2\uparrow} \hat{c}_{2\downarrow}^{\dag} \hat{c}_{2\downarrow},\nonumber \\
\hat{W} &= (W_1 - i W_2) \hat{c}_{1\u}^{\dag} \hat{c}_{1\d}^{\dag} \hat{c}_{2\d} \hat{c}_{2\u} + (W_1 + i W_2) \hat{c}_{2\u}^{\dag} \hat{c}_{2\d}^{\dag} \hat{c}_{1\d} \hat{c}_{1\u}. \nonumber
\end{align}
The spin-summed one-matrix in this model is simply a Hermitian $2\times 2$ matrix $\gamma = I + \vec{\gamma}\cdot\vec{\sigma}$, where $\vec{\gamma}= A(\sin \theta \cos \varphi, \sin \theta \sin \varphi, \cos \theta)$ is similar to a pseudospin vector with Bloch sphere angles $\theta$ and $\varphi$, but its modulus $A$ is less than $1$ due to correlations.  The natural orbitals are chosen to be $\phi_a = (\cos(\theta/2) e^{-i\varphi/2}, \sin(\theta/2) e^{i\varphi/2})^T$ and $\phi_b = (-\sin(\theta/2) e^{-i\varphi/2}, \cos(\theta/2) e^{i\varphi/2})^T$.  The corresponding occupation numbers are $n_a = 1 + A$ and $n_b = 1 - A$.  Equations~(\ref{eqn:orb}) and (\ref{eqn:occnum}) become
\begin{align}
i \dot{\phi}_k &= \big( \frac{1}{2} \vec{V}\cdot\vec{\sigma} + v_{ee} \big) \phi_k 
\label{eqn:orb:model} \\
\dot{A} &= \frac{1}{2} (u_{aa}-u_{bb}) {.} \label{eqn:occnum:model}
\end{align}
In Eq.~(\ref{eqn:orb:model}), $v_{ee}$ is the contribution of $\hat{U}$ and $\hat{W}$ to the effective single-particle Hamiltonian.  In the natural orbital basis, its off-diagonal elements are $v_{ee,ab} = v_{ee,ba}^* = -i u_{ab}/2A$, while its diagonal elements are indeterminate because the phases of the natural orbitals are undefined.  Direct calculation with $|\Psi^{LS}\rangle$ in Eq.~(\ref{eqn:psi:LS}) gives
\begin{align}
u_{ab} &= -i \overline{U} (1+B e^{-i2\zeta}) \sin \theta \cos \theta -i \Delta U A \sin \theta \nonumber \\
&+i |W| (1+B e^{-i2\zeta}) \sin \theta \cos \theta \cos(2\varphi-\omega) \nonumber \\
&- |W| (1-B e^{-i2\zeta}) \sin \theta \sin(2\varphi-\omega) {,}
\end{align}
where $B=\sqrt{1-A^2}$, $|W| e^{-i\omega} = W_1-i W_2$, $\overline{U}=(U_1+U_2)/2$, $\Delta U = (U_1-U_2)/2$ and $\zeta = \zeta_a-\zeta_b$ is the relative two-matrix phase.  The diagonal elements of $u$ enter in Eq.~(\ref{eqn:occnum:model}).  We have $u_{aa} = -u_{bb}$ and
\begin{align}
u_{aa} &= -\overline{U} B \sin^2 \theta \sin2\zeta - 2|W| B \cos \theta \sin(2\varphi-\omega) \nonumber \\
&\times \cos 2\zeta - |W| B (1+\cos^2 \theta) \cos(2\varphi-\omega) \sin 2\zeta{.} \nonumber 
\end{align}
The dynamical equation for $\zeta$ can be derived from the stationarity of the action $S[\Psi]=\int_0^T dt \langle \Psi | \hat{H} - i\partial_t |\Psi\rangle$ with respect to $A$. 
We find
\begin{align}
\dot{\zeta} &= \frac{1}{A}\vec{V}\cdot \vec{\gamma} + \frac{\overline{U}}{2} \frac{A}{B} \sin^2 \theta \cos2\zeta + \Delta U \cos \theta + \frac{|W|}{2} \frac{A}{B} \nonumber \\
&\times (1+\cos^2\theta) \cos(2\varphi-\omega) \cos2\zeta - |W| \frac{A}{B} \cos \theta  \nonumber \\
&\times \sin(2\varphi-\omega) \sin2\zeta - \dot{\varphi} \cos \theta {.} \label{eqn:zeta:1}
\end{align}
Here, we have used $\langle \Psi | \hat{H} |\Psi\rangle = \vec{V}\cdot\vec{\gamma} + U[\Psi] + W[\Psi]$ with
\begin{align}
U[\Psi] &= \frac{\overline{U}}{2}(1+\cos^2\theta) - \frac{\overline{U}}{2} B \sin^2 \theta \cos2\zeta + \Delta U A \cos \theta {,} \nonumber \\
W[\Psi] &= \frac{|W|}{2} \sin^2 \theta \cos(2\varphi-\omega) + |W| B \cos\theta \sin(2\varphi-\omega) \nonumber \\
&\times\sin2\zeta - \frac{|W|}{2} B (1+\cos^2\theta) \cos(2\varphi-\omega) \cos2\zeta  \nonumber 
\end{align}
and $i\int_0^T dt \langle \Psi | \dot{\Psi} \rangle = \int_0^T dt (\dot{\mu} + A \dot{\varphi} \cos \theta + A\dot{\zeta})$.  
 
\begin{figure}[t!]
\includegraphics[width=1.0\columnwidth]{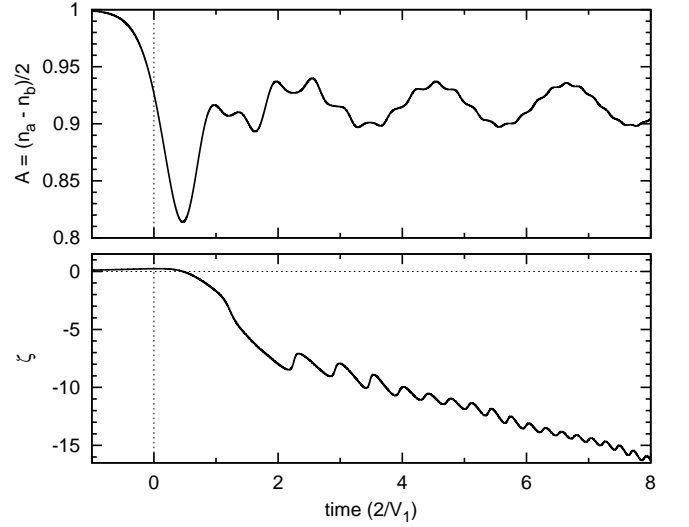}
\caption{\label{fig:AZeta} Time dependence of $A$ and $\zeta$ for $U_1=U_2=3/2$, $|W|=0$, and driving potential $\vec{V} = (-2,0,4t)$.}
\end{figure}  

As Eqs.~(\ref{eqn:orb:model}), (\ref{eqn:occnum:model}) and (\ref{eqn:zeta:1}) describe all of the degrees of freedom of $|\Psi^{LS}\rangle$ in Eq.~(\ref{eqn:psi:LS}), they are equivalent to the Schr\"odinger equation.  In RDMFT, $\zeta$ is interpreted as a functional of $\gamma(t)$.  If the mapping $\vec{V}(t)\rightarrow \vec{\gamma}(t)$ is invertible for a given initial state, the memory-dependent functional $\zeta([\gamma],t)$ is given uniquely by the solution of Eq.~(\ref{eqn:zeta:1}).  In Fig.~\ref{fig:AZeta}, $A$ and $\zeta$ are plotted for a linear-time driving potential $\vec{V} = (-2,0,4t)$.  The initial state is the instantaneous ground state at $t_0=-4$.  Although in numerical simulations it is not possible to specify the initial condition at $t=-\infty$ as in the LZ model, we chose an initial time early enough so that the td $A$ and $\zeta$ are numerically converged with respect to the limit $t_0 \rightarrow -\infty$.  In the limit $t \rightarrow \infty$, $A$ displays persistent oscillations with frequency $2U$.  This is surprising because in the long-time limit, where the system is a nonstationary state, one expects the oscillations to have the frequencies $\Omega_{ij} \equiv E_i-E_j$, where $E_i$ are the adiabatic energy levels.  The components of $\vec{\gamma}$ do indeed oscillate with the frequencies $\Omega_{ij}$.  Due to the divergence of the linear-time driving potential $\vec{V}$, the $\Omega_{ij}$ diverge in the limit $t\rightarrow \infty$ ($E_1 \sim -V_3+U$, $E_2 \sim 0$ and $E_3 \sim V_3+U$).  In contrast, $A$, which describes the occupation number degrees of freedom, oscillates with the \textit{constant} frequency $\Omega_{32}-\Omega_{21} = 2U$.  This expression demonstrates that all three states are participating in the oscillations; they have no analog in two-state systems.  The emergence of the Hubbard energy $U$ is a consequence of the fact that the occupation numbers are driven exclusively by the internal correlation of the system.  One can expect to see \textit{correlation-induced} oscillations in $n_k$ whenever a many-body system traverses a correlated avoided crossing.

We also observe that $\zeta$ decreases linearly in the limit $t\rightarrow \infty$.  The asymptotic slope, $-U$, is exactly half the frequency of the oscillations in $A$.  The relationship between $A$ and $\zeta$ in the long-time regime will be discussed further in Sec.~\ref{ssec:long}.  Superimposed on the linear dependence are nonlinear oscillations with a ``sawtooth'' pattern.  The rapid jumps in $\zeta$ come from the term $\dot{\varphi}\cos\theta$ in Eq.~(\ref{eqn:zeta:1}).  To understand this, consider the dynamics of $\vec{\gamma}$ in three-dimensional space.  The vector $\vec{\gamma}$ begins at the north pole ($\theta=0$) of the Bloch sphere at $t=-\infty$.  Then, it follows adiabatically the driving vector $-\vec{V}$ as it rotates toward the south pole.  Although $-\vec{V}/|\vec{V}|$ approaches the south pole in the limit $t\rightarrow \infty$, $\vec{\gamma}$ does not.  Instead, due to non\-adiabatic transitions, it precesses continuously around the south pole with constant $\gamma_{3,\infty}\equiv \gamma_3(\infty)$.  However, before it reaches this asymptotic behavior, there is an interval of time during which, periodically, $\vec{\gamma}$ passes close to the south pole (see Fig.~1 in Ref.~\onlinecite{requist2010a}).  For each such time, $\dot{\varphi}$ is strongly peaked, thus inducing jumps in $\zeta$.  

The phases $\zeta_k$ can be given a geometric interpretation.  Consider a cyclic evolution on the time interval $[0,T]$.  A cyclic evolution is one for which $|\Psi(T)\rangle$ differs from $|\Psi(0)\rangle$ by only an overall phase, i.e. $|\Psi(T)\rangle=e^{-i\nu}|\Psi(0)\rangle$.  The geometric phase \cite{aharonov1987} for the general two-electron spin-singlet state in Eq.~\ref{eqn:psi:LS} is
\begin{align}
i\int_0^T dt \langle \xi | \dot{\xi} \rangle = \sum_k \oint (i n_k \left<\phi_k|d\phi_k\right> + n_k d\zeta_k),  \label{eqn:geometric:phase}
\end{align}
where $|\xi(t)\rangle=e^{i\mu(t)}|\Psi(t)\rangle$ and $\mu(t)$ is any real-valued function for which $\mu(T)-\mu(0)=\nu$.  The first term in Eq.~\ref{eqn:geometric:phase} is the geometric phase associated with the natural orbitals.  The second term shows that the time-evolving $\zeta_k$ give a contribution to the total geometric phase above and beyond the natural orbital contribution.  In the present model, the geometric phase simplifies to $\oint (A \cos\theta d\varphi + A d\zeta)$.  In the noninteracting case, this becomes $\oint \cos\theta d\varphi$, which is just the familiar result for the pseudospin of a two-level system.  Interactions modify the noninteracting geometric phase in two ways.  First, the factor $A$, which is between $0$ and $1$, reduces the natural orbital contribution.  From its definition, we have $A=(n_a-n_b)/2$, where $n_k = n_{k\uparrow}+n_{k\downarrow}$ and $n_a\geq n_b$.  $A=1$ in the noninteracting case, and it decreases as the correlation of the state increases.  Second, interactions introduce the additional term $\sum_k \oint n_k d\zeta_k = \oint A d\zeta$.  This term vanishes in the noninteracting case because for $n_k=const$, $\oint n_k d\zeta_k = n_k \oint d\zeta_k = 0$.  Interactions have a similar effect on the gs Berry phase. 

\begin{figure}[t!]
\includegraphics[width=1.0\columnwidth]{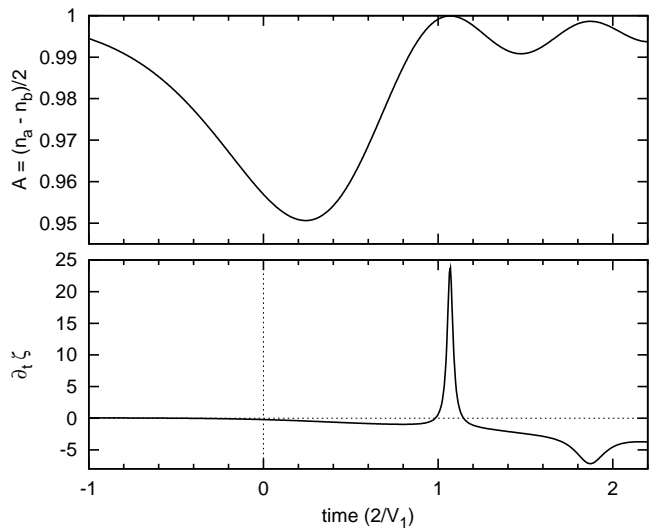}
\caption{\label{fig:resonance} Example of a resonance in $\zeta$ when $n_k$ approaches $1$.  Parameters are $U_1=U_2=1$, $|W|=0$, and $\vec{V} = (-2,0,2t)$.}
\end{figure}  
 
It is important to remember that $\zeta$ is a \textit{relative} phase that depends on our choice of td natural orbital phases.  Redefining the natural orbital phases so as to satisfy the parallel transport condition $\mathrm{Im}\langle \phi_k | \dot{\phi}_k \rangle=0$ redefines $\zeta$ according to $\dot{\zeta} \rightarrow \dot{\zeta} - \dot{\varphi} \cos\theta$.  Then, the geometric phase in Eq.~(\ref{eqn:geometric:phase}) becomes $i\int_0^T dt \langle \xi | \dot{\xi} \rangle = \sum_k \oint n_k d\zeta_k = \oint A d\zeta$.  

There is a remarkable aspect of the $\zeta_k$ worth discussing.  When one of the $n_k$ approaches one of the boundaries of the interval $[0,1]$, the phase $\zeta_k$ undergoes a resonance, jumping by $\pi/2$.  The effect of this resonance is to change the sign of $\dot{n}_k$, thereby keeping the occupation numbers in the allowed interval $[0,1]$.  The upper bound is a consequence of the Pauli exclusion principle.

The width of the resonance depends on how close $n_k$ comes to the boundary.  Figure~\ref{fig:resonance} shows $A$ and $\dot{\zeta}$ in an interval where $A$ approaches close to $1$, i.e. $n_{a\uparrow}=n_{a\downarrow}\rightarrow 1$ and $n_{b\uparrow}=n_{b\downarrow} \rightarrow 0$.  We can obtain a \textit{universal} equation for $\zeta$ near the resonance by keeping only the leading terms in the limit $A\rightarrow 1$, which is equivalent to $B\rightarrow 0$.  From Eqs.~(\ref{eqn:occnum:model}) and (\ref{eqn:zeta:1}), we find
\begin{align}
\ddot{\zeta} + 2\f{\dot{B}}{B} \dot{\zeta} = 0. \label{eqn:zeta:universal}
\end{align}
Integration gives $\dot{\zeta} = const/B^2$, and further integration gives the memory-dependent functional $\zeta([\gamma],t)$.  Choosing the case $|W|=0$ for simplicity, the explicit solution of Eq.~(\ref{eqn:zeta:universal}) is
\begin{align}
\zeta = \f{1}{2} \tan^{-1}[2 \alpha(t-t_c)].
\end{align}
This is the formula for a resonance of width $\alpha^{-1}$ centered at $t=t_c$.  We expect a similar resonance phenomenon will occur in $N$-electron systems at the boundary $n_k=1$.  The boundary $n_k=0$ might be more subtle, as it is an accumulation point of the spectrum of $\gamma$.

\subsection{\label{ssec:long} Long-time regime}

We consider the long-time limit of the model when it starts in the ground state at $t=-\infty$ and experiences the external driving $\vec{V} = (V_1, 0, t/\tau)$.  For simplicity, we set $U\equiv U_1=U_2$ and $|W|=0$.  The exact asymptotic behavior of $A(t)$ in the limit $t\rightarrow \infty$ is 
\begin{align}
A(t) = \sqrt{\overline{A}^2 + \Delta^2 \cos\big[2Ut - (\Theta_{32}-\Theta_{21})\big]}, \label{eqn:A:sq}
\end{align}
where $\overline{A}^2 \equiv \gamma_{3,\infty}^2 + 2 |c_2|^2 (|c_1|^2 + |c_3|^2)$, $\Delta^2\equiv 4 |c_1| |c_2|^2 |c_3|$, $\Theta_{ij} \equiv \mathrm{Arg}(c_i/c_j)$ and $c_i$ are the coefficients in the expansion $|\Psi\rangle = \sum_{i=1}^3 c_i \mathrm{exp}[-i\int dt (E_i -i \langle \Phi_i | \partial_t \Phi_i\rangle)] |\Phi_i\rangle$ over the adiabatic eigenstates $|\Phi_i\rangle$.  Since the pseudomagnetic field $\vec{V}$ diverges as $t\rightarrow \infty$, the pseudospin-like vector $\vec{\gamma}$ precesses more and more rapidly around the south pole of the Bloch sphere.  As $t\rightarrow \infty$, the azimuthal angle $\varphi$ grows quadratically in time, i.e., $\varphi \sim t^2/2\tau$.  Hence, any terms in the dynamical equations containing periodic functions of $\varphi$ quickly average to zero.  Dropping such terms in Eqs.~(\ref{eqn:occnum:model}) and (\ref{eqn:zeta:1}), we find 
\begin{align}
\dot{A} &= -U B \sin^2\theta \sin 2\zeta \label{eqn:A:asym}\\
\dot{\zeta} &= -U \frac{1+B\cos2\zeta}{A} \cos^2\theta + \frac{U}{2} \frac{A}{B} \sin^2 \theta \cos 2\zeta. \label{eqn:zeta:asym}
\end{align}
Although $\theta$ appears in these equations, it can be eliminated in favor of $A$ by means of the asymptotic relationship $\gamma_{3,\infty} = A \cos\theta$.  Therefore, asymptotically the equations for $A$ and $\zeta$ decouple from those for the orbital variables $\theta$ and $\varphi$.  Equations~(\ref{eqn:A:asym}) and (\ref{eqn:zeta:asym}) can be expressed in the form of Hamilton's canonical equations, 
\begin{align}
\dot{A} &= - \frac{\partial V_{ee,\infty}}{\partial \zeta} \label{eqn:Hamilton:A}\\
\dot{\zeta} &= \frac{\partial V_{ee,\infty}}{\partial A}, \label{eqn:Hamilton:zeta}
\end{align}
where $V_{ee,\infty} \equiv \lim_{t\rightarrow \infty} V_{ee}[\Psi]$ acts as an effective Hamiltonian and $A$ and $\zeta$ appear as canonically conjugate action-angle variables.  Equations~(\ref{eqn:Hamilton:A}) and (\ref{eqn:Hamilton:zeta}) are integrable and $V_{ee,\infty}$ is a constant of the motion.  Since $V_{ee,\infty}$ can be expressed as a function of $A$, $\zeta$ and $\gamma_{3,\infty}$, there is the following \textit{instantaneous} relationship between $A$ and $\zeta$ in the long-time limit:
\begin{align}
\cos 2\zeta = \frac{2A^2\big(1-\frac{V_{ee,\infty}}{U}\big) - (A^2 - \gamma_{3,\infty}^2)}{ \sqrt{1-A^2} (A^2-\gamma_{3,\infty}^2)}. \label{eqn:cos:zeta}
\end{align}
Therefore, the $\gamma$ dependence of $\zeta([\gamma],t)$ separates into two distinct types: (i) an ultra-local (instantaneous) dependence on $A(t)$ and (ii) an ultra-nonlocal dependence on $\gamma(t)$ near $t=0$ that enters only through the constant $\gamma_{3,\infty}$.  The dependence on $\gamma_{3,\infty}$ contains information about the non\-adiabatic transitions that occurred near $t=0$.  In the long-time regime, it can be viewed as initial-state dependence specified after the avoided crossings are complete.  Memory dependence becomes simple because nonadiabatic transitions are no longer occurring in the long-time regime.  Equations~(\ref{eqn:A:asym}-\ref{eqn:Hamilton:zeta}) suggest that memory dependence among certain degrees of freedom will be weaker when their mutual dynamics are nearly integrable and nearly decoupled from other degrees of freedom.

\subsection{\label{ssec:short} Short-time regime}

We now study $\zeta$ shortly after an external driving potential is turned on for a system in the ground state at finite $t=t_0$.  The driving potential is taken to have the form $\vec{V} = (V_1, V_2, V_3(t))$, where $V_1$ and $V_2$ are constant and $V_3(t)$ is an arbitrary continuous function for which $\dot{V}_3(t_0)\neq 0$.  This is analogous to a td local external potential in a continuous system.  The stationary conditions imply $\dot{\gamma}(t_0) = 0$ and $\dot{\zeta}(t_0)=0$. The lowest nonvanishing time derivative at the initial time is $\ddot{\varphi}(t_0) = \dot{V}_3(t_0)$.  This induces nonvanishing third time derivatives for $\theta$, $\varphi$, and $\zeta$.  The lowest nonvanishing time derivative of $A$ is of \textit{fourth} order.  The relative phase and the occupation numbers change even more slowly if $|W|=0$.  In this case, the lowest nonvanishing time derivative of $A$ is the fifth time derivative, and the changes proceed via $\dot{V}_3 \rightarrow \ddot{\varphi} \rightarrow \dddot{\theta} \rightarrow \zeta^{(4)} \rightarrow A^{(5)}$.  This means that the external driving induces first a current, the current leads to changes in the density, the changes in the density cause changes in $\zeta$, which, finally, induce changes in the occupation numbers.

\section{\label{sec:conclusions} Conclusions}

The occupation numbers are important degrees of freedom that provide information about the correlation of a many-body state.  Their dynamics are determined by the equation of motion for the one-body reduced density matrix.  This equation of motion contains the two-body reduced density matrix $\Gamma(t)$, and we have proved that approximating $\Gamma(t)$ by the adiabatic extension of the ground-state functional $\Gamma[\gamma]$ does not generate time-dependent occupation numbers.  For two-electron systems, this deficiency can be explicitly traced to the fact that the ground-state $\Gamma[\gamma]$ lacks dynamical relative phases $\zeta_k$ present in the exact $\Gamma(t)$.  Physically, the $\zeta_k$ describe nonadiabatic interaction effects.  The variables $n_k$ and $\zeta_k$ are canonically conjugate variables that give a contribution to the geometric phase above and beyond the contribution from the parallel transport of the natural orbitals.  Additionally, they have the important function of maintaining compliance with the Pauli exclusion principle.
 
We derived an exact differential equation for the $\zeta_k$ in a generalization of the Landau-Zener model that includes interactions.  Introducing two-body interactions splits the single Landau-Zener avoided crossing of two levels into a set of three pairwise avoided crossings among three levels.  After the system traverses these \textit{correlated} avoided crossings, the occupation numbers display oscillations whose frequency is determined by the Hubbard interaction $U$.  These correlation-induced oscillations, which depend on the participation of all three states in an essential way, are an observable effect of interactions in quantum many-body dynamics.

How to account for memory dependence in time-dependent density functional theories, especially in real time, is an important unsolved problem.  Equation~(\ref{eqn:zeta:1}) gives an explicit example of a differential equation whose solution determines the memory-dependent functional $\zeta([\gamma],t)$ and thereby the complete functional $\Gamma([\gamma],t)$.  We have found that the $\gamma$ dependence assumes a simple form in the long-time regime, comprising both  an instantaneous dependence and an ultra-nonlocal dependence representing \textit{past} nonadiabatic transitions.  Memory dependence also simplifies within the so-called independent crossing approximation, where each avoided crossing is treated as independent of the others.  At this level of approximation, memory dependence enters via dynamic and geometric phases and the amplitudes $c_i$ of the adiabatic eigenstates \cite{requist2010a}; the $c_i$, which are assumed to be constant for all states except those participating in a given avoided crossing, retain the memory of past nonadiabatic transitions.  Conversely, the exact memory dependence is nontrivial while nonadiabatic transitions are occurring, e.g., while the system is traversing an avoided crossing.  

Two adiabatic energy levels that undergo an avoided crossing in real time generally intersect at certain points in the complex plane of time.  The functional dependence of $\Gamma([\gamma],t)$ might simplify at such times, and indeed, the contour integral in the Dykhne formula suggests that in the adiabatic regime memory dependence might be easier to handle in complex time.  We also note that Eq.~(\ref{eqn:zeta:1}), viewed as a differential equation in the complex plane of time, has singularities when $n_a=n_b$.  
In the interacting Landau-Zener model, we have found that the instantaneous ground-state $n_a$ and $n_b$ have intersections in the neighborhood of the intersections of $E_1$ and $E_2$ \cite{requist2011x}.  
Hence, the occupation numbers contain useful information about the avoided crossings of the adiabatic energy levels.  

The interacting Landau-Zener model studied here might provide insight into the non\-adiabatic dynamics of many-body systems.  In situations where a pair of natural orbitals are strongly coupled and the others can be treated as an approximately inert background, it is possible to map the subspace dynamics onto the interacting Landau-Zener model.  The model parameters $\vec{V}$, $U$ and $W$, representing the effective ``screening'' of the other degrees of freedom, will be time dependent.  It might be possible to devise a memory-dependent functional based on Eq.~(\ref{eqn:zeta:1}) and the assumption that such pairwise interactions of the natural orbitals can be treated one at a time.

\begin{acknowledgments}
This work was supported by the Deutsche Forschungsgemeinshaft (Grant No. PA 516/7-1). 
\end{acknowledgments}

\bibliography{bibliography2010c}

\begin{thebibliography}{19}
\expandafter\ifx\csname natexlab\endcsname\relax\def\natexlab#1{#1}\fi
\expandafter\ifx\csname bibnamefont\endcsname\relax
  \def\bibnamefont#1{#1}\fi
\expandafter\ifx\csname bibfnamefont\endcsname\relax
  \def\bibfnamefont#1{#1}\fi
\expandafter\ifx\csname citenamefont\endcsname\relax
  \def\citenamefont#1{#1}\fi
\expandafter\ifx\csname url\endcsname\relax
  \def\url#1{\texttt{#1}}\fi
\expandafter\ifx\csname urlprefix\endcsname\relax\def\urlprefix{URL }\fi
\providecommand{\bibinfo}[2]{#2}
\providecommand{\eprint}[2][]{\url{#2}}

\bibitem[{\citenamefont{Gilbert}(1975)}]{gilbert1975}
\bibinfo{author}{\bibfnamefont{T.~L.} \bibnamefont{Gilbert}},
  \bibinfo{journal}{Phys. Rev. B} \textbf{\bibinfo{volume}{12}},
  \bibinfo{pages}{2111} (\bibinfo{year}{1975}).

\bibitem[{\citenamefont{Kohn and Sham}(1965)}]{kohn1965}
\bibinfo{author}{\bibfnamefont{W.}~\bibnamefont{Kohn}} \bibnamefont{and}
  \bibinfo{author}{\bibfnamefont{L.~J.} \bibnamefont{Sham}},
  \bibinfo{journal}{Phys. Rev.} \textbf{\bibinfo{volume}{140}},
  \bibinfo{pages}{A1133} (\bibinfo{year}{1965}).

\bibitem[{\citenamefont{L\"owdin}(1955)}]{loewdin1955}
\bibinfo{author}{\bibfnamefont{P.~O.} \bibnamefont{L\"owdin}},
  \bibinfo{journal}{Phys. Rev.} \textbf{\bibinfo{volume}{97}},
  \bibinfo{pages}{1474} (\bibinfo{year}{1955}).

\bibitem[{\citenamefont{Appel}(2007)}]{appel2007}
\bibinfo{author}{\bibfnamefont{H.}~\bibnamefont{Appel}}, Ph.D. thesis,
  \bibinfo{school}{Freie Universitaet Berlin} (\bibinfo{year}{2007}),
  \urlprefix\url{<http://www.diss.fu-berlin.de/2007/481/>}.

\bibitem[{\citenamefont{Appel and Gross}(2010)}]{appel2010}
\bibinfo{author}{\bibfnamefont{H.}~\bibnamefont{Appel}} \bibnamefont{and}
  \bibinfo{author}{\bibfnamefont{E.~K.~U.} \bibnamefont{Gross}},
  \bibinfo{journal}{Eur. Phys. Lett.} \textbf{\bibinfo{volume}{92}},
  \bibinfo{pages}{23001} (\bibinfo{year}{2010}).

\bibitem[{\citenamefont{Giesbertz et~al.}(2008)\citenamefont{Giesbertz,
  Baerends, and Gritsenko}}]{giesbertz2008}
\bibinfo{author}{\bibfnamefont{K.~J.~H.} \bibnamefont{Giesbertz}},
  \bibinfo{author}{\bibfnamefont{E.~J.} \bibnamefont{Baerends}},
  \bibnamefont{and} \bibinfo{author}{\bibfnamefont{O.~V.}
  \bibnamefont{Gritsenko}}, \bibinfo{journal}{Phys. Rev. Lett.}
  \textbf{\bibinfo{volume}{101}}, \bibinfo{pages}{033004}
  (\bibinfo{year}{2008}).

\bibitem[{\citenamefont{Giesbertz et~al.}(2009)\citenamefont{Giesbertz, Pernal,
  Gritsenko, and Baerends}}]{giesbertz2009}
\bibinfo{author}{\bibfnamefont{K.~J.~H.} \bibnamefont{Giesbertz}},
  \bibinfo{author}{\bibfnamefont{K.}~\bibnamefont{Pernal}},
  \bibinfo{author}{\bibfnamefont{O.~V.} \bibnamefont{Gritsenko}},
  \bibnamefont{and} \bibinfo{author}{\bibfnamefont{E.~J.}
  \bibnamefont{Baerends}}, \bibinfo{journal}{J. Phys. Chem.}
  \textbf{\bibinfo{volume}{130}}, \bibinfo{pages}{114104}
  (\bibinfo{year}{2009}).

\bibitem[{\citenamefont{Pernal et~al.}(2007{\natexlab{a}})\citenamefont{Pernal,
  Gritsenko, and Baerends}}]{pernal2007a}
\bibinfo{author}{\bibfnamefont{K.}~\bibnamefont{Pernal}},
  \bibinfo{author}{\bibfnamefont{O.}~\bibnamefont{Gritsenko}},
  \bibnamefont{and} \bibinfo{author}{\bibfnamefont{E.~J.}
  \bibnamefont{Baerends}}, \bibinfo{journal}{Phys. Rev. A}
  \textbf{\bibinfo{volume}{75}}, \bibinfo{pages}{012506}
  (\bibinfo{year}{2007}{\natexlab{a}}).

\bibitem[{\citenamefont{Rajam et~al.}(2009)\citenamefont{Rajam, Hessler, Gaun,
  and Maitra}}]{rajam2009}
\bibinfo{author}{\bibfnamefont{A.~K.} \bibnamefont{Rajam}},
  \bibinfo{author}{\bibfnamefont{P.}~\bibnamefont{Hessler}},
  \bibinfo{author}{\bibfnamefont{C.}~\bibnamefont{Gaun}}, \bibnamefont{and}
  \bibinfo{author}{\bibfnamefont{N.~T.} \bibnamefont{Maitra}},
  \bibinfo{journal}{J. Molec. Struct.} \textbf{\bibinfo{volume}{914}},
  \bibinfo{pages}{30} (\bibinfo{year}{2009}).

\bibitem[{\citenamefont{Runge and Gross}(1984)}]{runge1984}
\bibinfo{author}{\bibfnamefont{E.}~\bibnamefont{Runge}} \bibnamefont{and}
  \bibinfo{author}{\bibfnamefont{E.~K.~U.} \bibnamefont{Gross}},
  \bibinfo{journal}{Phys. Rev. Lett.} \textbf{\bibinfo{volume}{52}},
  \bibinfo{pages}{997} (\bibinfo{year}{1984}).

\bibitem[{\citenamefont{Ghosh and Dhara}(1988)}]{ghosh1988}
\bibinfo{author}{\bibfnamefont{S.~K.} \bibnamefont{Ghosh}} \bibnamefont{and}
  \bibinfo{author}{\bibfnamefont{A.~K.} \bibnamefont{Dhara}},
  \bibinfo{journal}{Phys. Rev. A} \textbf{\bibinfo{volume}{38}},
  \bibinfo{pages}{1149} (\bibinfo{year}{1988}).

\bibitem[{\citenamefont{Vignale}(2004)}]{vignale2004}
\bibinfo{author}{\bibfnamefont{G.}~\bibnamefont{Vignale}},
  \bibinfo{journal}{Phys. Rev. B} \textbf{\bibinfo{volume}{70}},
  \bibinfo{pages}{201102(R)} (\bibinfo{year}{2004}).

\bibitem[{\citenamefont{Requist and
  Pankratov}(2010{\natexlab{a}})}]{requist2010a}
\bibinfo{author}{\bibfnamefont{R.}~\bibnamefont{Requist}} \bibnamefont{and}
  \bibinfo{author}{\bibfnamefont{O.}~\bibnamefont{Pankratov}},
  \bibinfo{journal}{Phys. Rev. A} \textbf{\bibinfo{volume}{81}},
  \bibinfo{pages}{042519} (\bibinfo{year}{2010}{\natexlab{a}}).

\bibitem[{\citenamefont{Pernal et~al.}(2007{\natexlab{b}})\citenamefont{Pernal,
  Giesbertz, Gritsenko, and Baerends}}]{pernal2007b}
\bibinfo{author}{\bibfnamefont{K.}~\bibnamefont{Pernal}},
  \bibinfo{author}{\bibfnamefont{K.}~\bibnamefont{Giesbertz}},
  \bibinfo{author}{\bibfnamefont{O.}~\bibnamefont{Gritsenko}},
  \bibnamefont{and} \bibinfo{author}{\bibfnamefont{E.~J.}
  \bibnamefont{Baerends}}, \bibinfo{journal}{J. Chem. Phys.}
  \textbf{\bibinfo{volume}{127}}, \bibinfo{pages}{214101}
  (\bibinfo{year}{2007}{\natexlab{b}}).

\bibitem[{\citenamefont{Giesbertz et~al.}(2010)\citenamefont{Giesbertz,
  Gritsenko, and Baerends}}]{giesbertz2010b}
\bibinfo{author}{\bibfnamefont{K.~J.~H.} \bibnamefont{Giesbertz}},
  \bibinfo{author}{\bibfnamefont{O.~V.} \bibnamefont{Gritsenko}},
  \bibnamefont{and} \bibinfo{author}{\bibfnamefont{E.~J.}
  \bibnamefont{Baerends}}, \bibinfo{journal}{Phys. Rev. Lett.}
  \textbf{\bibinfo{volume}{105}}, \bibinfo{pages}{013002}
  (\bibinfo{year}{2010}).

\bibitem[{\citenamefont{Requist and
  Pankratov}(2010{\natexlab{b}})}]{requist2010b}
\bibinfo{author}{\bibfnamefont{R.}~\bibnamefont{Requist}} \bibnamefont{and}
  \bibinfo{author}{\bibfnamefont{O.}~\bibnamefont{Pankratov}},
  \bibinfo{howpublished}{arxiv:1011.1482} (\bibinfo{year}{2010}{\natexlab{b}}).

\bibitem[{\citenamefont{L\"owdin and Shull}(1956)}]{loewdin1956}
\bibinfo{author}{\bibfnamefont{P.~O.} \bibnamefont{L\"owdin}} \bibnamefont{and}
  \bibinfo{author}{\bibfnamefont{H.}~\bibnamefont{Shull}},
  \bibinfo{journal}{Phys. Rev.} \textbf{\bibinfo{volume}{101}},
  \bibinfo{pages}{1730} (\bibinfo{year}{1956}).

\bibitem[{\citenamefont{Aharonov and Anandan}(1987)}]{aharonov1987}
\bibinfo{author}{\bibfnamefont{Y.}~\bibnamefont{Aharonov}} \bibnamefont{and}
  \bibinfo{author}{\bibfnamefont{J.}~\bibnamefont{Anandan}},
  \bibinfo{journal}{Phys. Rev. Lett.} \textbf{\bibinfo{volume}{58}},
  \bibinfo{pages}{1593} (\bibinfo{year}{1987}).

\bibitem[{\citenamefont{Requist and Pankratov}()}]{requist2011x}
\bibinfo{author}{\bibfnamefont{R.}~\bibnamefont{Requist}} \bibnamefont{and}
  \bibinfo{author}{\bibfnamefont{O.}~\bibnamefont{Pankratov}},
  \bibinfo{howpublished}{unpublished.}

\end{thebibliography}

\end{document}